\documentclass[11pt,twocolumn]{IEEEtran}

\usepackage{balance}
\usepackage{subfigure}
\usepackage{amssymb}
\usepackage{amsmath}
\usepackage{graphicx}
\usepackage{color}
\usepackage{multirow}
\usepackage{amssymb}
\usepackage{pifont}
\usepackage{tikz,pgfplots} 
\usepackage{booktabs, multicol, multirow}
\usepackage[acronym,nomain]{glossaries}

\usepackage[inline]{enumitem}
\usepackage{algorithm}
\usepackage{algorithmicx}
\usepackage{algpseudocode}
\usepackage{amsmath,amsfonts,amssymb}
\usepackage[small,bf,hypcap=false]{caption}
\usepackage{subfigure}
\usepackage{grffile}
\usepackage{tabularx}
\usepackage{colortbl}
\usepackage{hhline}
\usepackage{hyperref}
\usepackage[flushleft]{threeparttable}
\usepackage{booktabs,fixltx2e}
\usepackage{balance}
\usepackage{boldline}
\usepackage{rotating}
\usepackage{pdflscape}
\usepackage[many]{tcolorbox}
\usepackage{framed}
\usepackage{cite}
\colorlet{shadecolor}{blue!20}

\renewcommand{\mathbf}[1]{{\boldsymbol #1}}
\hypersetup{
	colorlinks   = true,
	citecolor    = blue,
	linkcolor    = blue,
}

\usepackage[english]{babel}
\usepackage{blindtext}

\newcommand{\cmark}{\ding{52}}%

\newcommand{\jack}[1]{\textcolor{red}{Jack says: #1}}
\newcommand{\minke}[1]{\textcolor{cyan}{Minke says: #1}} 

\begin{document}


\title{An Exploratory Study on Machine Learning Model Stores}

\author{\IEEEauthorblockN{Minke Xiu\IEEEauthorrefmark{1}, Zhen Ming (Jack) Jiang\IEEEauthorrefmark{2}, Bram Adams\IEEEauthorrefmark{3}}\\
\IEEEauthorblockA{
York University - Toronto, Canada\IEEEauthorrefmark{1}\IEEEauthorrefmark{2} - \{xmk233\IEEEauthorrefmark{1}, zmjiang\IEEEauthorrefmark{2}\}@eecs.yorku.ca
}~\\
\IEEEauthorblockA{
Polytechnique Montr\'{e}al, Montr\'{e}al, Canada\IEEEauthorrefmark{3} - bram.adams@polymtl.ca\IEEEauthorrefmark{3}
}~\\
}

\markboth{Accepted by IEEE Software}{2020}

\maketitle

\begin{abstract}
  Recent advances in Artificial Intelligence, especially in Machine Learning (ML), have brought applications previously considered as science fiction (e.g., virtual personal assistants and autonomous cars) into the reach of millions of everyday users. 
  Since modern ML technologies like deep learning require considerable technical expertise and resource to build custom models, reusing existing models trained by experts has become essential. This is why in the past year model stores have been introduced by several organizations (e.g., Amazon and Wolfram), which, inspired by mobile app stores, provide public access to pre-trained models and infrastructure to deploy/evaluate/re-train. 
  This paper conducts an exploratory study on three popular model stores that compares the information elements (features and policies) provided by them to those used by 
  two popular mobile app stores. We have found that the three stores share a common foundation (65\% of elements shared).
  There are also eight elements (e.g., cloud deployment and technical documentation) unique to model stores, as those are targeted towards software engineers instead of end-users (for mobile stores). Among the four elements unique to app stores, 
  two elements (review policy and permission) could also be applicable to model stores and have a critical impact on the safety and performance of ML applications. 
  The majority of the models from Wolfram and ModelDepot originate from published research prototypes. There are only a few similar offerings of ML models among the three stores. 
\end{abstract}


\section{Introduction}
\label{sec:introduction}

Artificial Intelligence is gaining rapid popularity in both research and practice, due to the recent advances in the research and development of machine learning (ML). Many ML applications (e.g., Tesla's autonomous vehicle and Apple's Siri) have already been used widely in people's everyday lives. McKinsey recently estimated that ML applications have the potential to create between \$3.5 and \$5.8 trillion in value annually \cite{MckinseyReport}.

Despite this rapid advance, the development of ML applications is quite different from conventional software applications~\cite{MeijerFSE2018Keynote}. For example, conventional software applications like e-commerce or mobile applications are developed based on specifications and carefully designed algorithms, whereas ML applications are mainly developed by feeding data into ML algorithms (a.k.a., the process of training ML models) followed by integrating the resulting models within a conventional application. Training ML models can be very costly, as it  requires running on expensive hardware equipment for a long period of time~\cite{ZophCVPR2018}. Therefore, similar to importing third party libraries for developing conventional systems, it is now a common practice to reuse pre-trained ML models for developing ML applications. In particular, Gartner has recently identified leveraging pre-trained ML models deployed as web services to be one of the top technology trends~\cite{GartnerTop10}. Popular frameworks like Tensorflow~\cite{TFModels} and PyTorch~\cite{PTVModels} provide high level API support to easily integrate a set of common ML models.

Since deploying and serving ML models is still a non-trivial task, which requires deep knowledge in ML and system adminstration~\cite{MoritzOSDI2018}, various organizations recently have ML models stores introduced to facilitate the distribution and retail of ML models to organizations/developers. 
Inspired by mobile app stores, model stores promise to fill the gap between highly specialized AI experts and software developers who are willing to integrate pre-trained ML models into their system, with or without customization (retraining).

However, what makes model stores unique in terms of the information elements (features and policies) presented about their products (models)?
What kind of models are offered in such stores? 
This paper conducts an exploratory study on three general-purpose model stores (AWS marketplace~\cite{awsmarketplaceformlai}, ModelDepot~\cite{modeldepot}, and Wolfram neural net repository~\cite{wolframneuralnetrepository}) to address the above two research questions. 
The contributions of this paper are:
\begin{itemize}

\item The first empirical study on ML model stores.

\item An overview on the current state-of-the-art of ML model stores. 

\item A discussion on identified practices and challenges on distributing and retailing pre-trained ML models.


\end{itemize}

\section{Background and Related Work}
\label{sec:relatedWorks}

\subsection{App Stores and Model Stores}
\label{sec:caseStudySetup}

Despite their difference in age, app and model stores provide platforms for developers to distribute and retail their products to their intended target audience (end users vs. organizations/developers).
App stores 
have been around for over ten years. Apple's App Store and Google Play, both started in 2008, are currently two of the most popular app stores. Each contains over two million apps.
These app stores 
include mobile apps and software applications for computers (e.g., Mac App Store) and tablets (e.g., Chromebook and iPad).
In contrast, the concept of ``model store'' is relatively new, with 
ModelDepot starting in  01/2018, Wolfram neural net repository in  06/2018, and AWS marketplace in  11/2018. For brevity, we will call these three model stores as ``ModelDepot'', ``Wolfram'', and ``AWS''. 

There are two types of model stores: (1) general purpose, and (2) specialized model stores. General purpose model stores (e.g., AWS) 
contain all sorts of ML models, whereas specialized model stores (e.g., Nuance AI market~\cite{nuanceaimarketplace}) only contain models from certain domains.
We focus on the general purpose model stores, since they target a much wider audience of organizations/developers and can provide a more representative view of SE practices on ML models. 

\subsection{Empirical Studies on Mobile App Stores}

  There is a large corpus of research on empirical studies of the mobile app stores (e.g., information elements~\cite{jansen2013defining}, user reviews~\cite{khalid2015mobile}, and update frequency~\cite{mcilroy2016fresh}) on understanding and improving the quality of the apps. 
  Since the information elements in app stores by now are widely understood, this paper focuses on an empirical comparison of app stores to the newly introduced model stores.
\section{RQ1: What kind of information elements do model stores provide?}
\label{sec:rq1}


This RQ compares 
(1) the information elements among three model stores; and (2) the information elements between model stores and app stores.
Collectively, we use the term ``product'' to refer to either an ML model or a mobile app. When referring to products from individual stores, we use the term ``models'' and ``apps'', respectively.


\subsection{Data Extraction}

For each of the three considered model stores, we used open coding~\cite{StolICSE2016} to label the structure of the webpages used to sell/provide models. Two coders separately split up each page into ``information elements'', sections that provide a specific functionality geared towards the store's clients. For example, a section can provide a description or the price of a product. 
We started with the two app stores and tried to rediscover the reported information elements from Jansen et al.~\cite{jansen2013defining}. 
Since two app stores may use terms differently, we manually merged the corresponding information elements among them.
Certain elements from that paper are in considerable detail (e.g., different revenue models), we merged those detailed elements under higher-level elements, where they apply. 
In the end, all information elements from~\cite{jansen2013defining} were found by our study. 
Furthermore, ten 
additional store elements were found by us concerning release notes and product permissions, as app stores kept evolving since 2013. 
A similar process was performed on model stores and new elements unique to the model stores were found. 
For all stores, we grouped related elements into larger dimensions (e.g., \texttt{user feedback}, \texttt{usage statistics}, \texttt{pricing} under the \textbf{Business} dimension). This process was conducted by the first two authors of this paper, and later verified by the third author to ensure correctness.

In the end, we have identified 
$26$ unique elements across six different dimensions among all the stores 
as shown in Table~\ref{tab:mergedTable}. Each row corresponds to one element, while a {\cmark} indicates the presence of that element in a given store. 

\newcolumntype{P}[1]{>{\centering\arraybackslash}p{#1}}

\begin{table*}[]
    \centering
    \caption{Comparing elements among the five mobile app and model stores. We use the term ``product'' to refer to both ``mobile apps'' and ``ML models''.}
    \label{tab:mergedTable}
    \begin{threeparttable}
    \begin{tabular}{P{2cm}P{2.5cm}P{1cm}P{1.5cm}P{1cm}P{1cm}P{1cm}p{5.5cm}}
    \toprule
    \multirow{2}{*}{\textbf{Dimension}} & \multirow{2}{*}{\textbf{Element}} & \multicolumn{3}{c}{\textbf{Model Store}} & \multicolumn{2}{c}{\textbf{App Store}} & \multicolumn{1}{c}{\textbf{Description}} \\ 

    & & (\textbf{AWS} & \textbf{ModelDepot} & \textbf{Wolfram}) & (\textbf{Apple} &\textbf{Google}) & \\ 

    \midrule


    \multirow{17}{2cm}{\centering Product Information} & Owner & \cmark & \cmark & \cmark & \cmark & \cmark & Developer information of this product. \\ 

    & Description & \cmark & \cmark & \cmark & \cmark & \cmark & The objectives and the functionalities of this product. \\ 

    & Demo &  & \cmark &  &  &  & A functionality provided for end users, so that they can try before buying/deploying the product. \\ 

    & Language &  &  &  & \cmark &  & Languages used in the user interface of this product. \\ 

    & Size &  & \cmark & \cmark & \cmark & \cmark & Size of the product in disk. \\ 

    & Version number & \cmark &  & \cmark & \cmark & \cmark &
    The version number of the current release.  \\ 

    & Permission &  &  &  &  & \cmark & The list of hardware/software resources needed from a user's device to properly run this product. \\ 

    & Age rating &  &  &  & \cmark & \cmark & Constraints about the user's age. \\ 



    \midrule

    \multirow{12}{2cm}{\centering Technical Documentation}  & User instruction & \cmark & \cmark & \cmark &  &  & Instructions on how to use this product. \\ 
    
    & Framework & \cmark & \cmark & \cmark &  &  & The underlying development framework for the ML algorithms used in this product.  \\ 

    & ML Algorithms & \cmark & \cmark & \cmark &  &  & The types of ML algorithms used in this product.  \\ 

    & Training set & \cmark & \cmark & \cmark &  &  & Datasets used for training the underlying ML algorithms.  \\ 

    & Performance & \cmark & \cmark & \cmark &  &  & The performance (e.g., precision, recall, and accuracy) of the underlying ML algorithms.  \\ 

    & Origin &  \cmark & \cmark & \cmark &  &  & Source of where the product originally came from (e.g., academic papers, open source products). \\ 

    & Release notes & \cmark &  &  & \cmark & \cmark & Information regarding the changes in the current version of the product.  \\ 



    \midrule

    \multirow{8}{2cm}{\centering Delivery} & Deployment instructions & \cmark & \cmark & \cmark &  &  & Instructions on how to deploy and configure the product. \\ 

    & Compatibility &  &  &  & \cmark & \cmark & Information on which platforms and versions are compatible with the product. \\ 

    & Local installation &  &  &  & \cmark & \cmark & Automated installation of the product to a user's device. \\ 

    & Cloud deployment & \cmark & \cmark & \cmark &  &  &
    Automatically deploying the product within the provider's cloud infrastructure. \\ 

    \midrule

    \multirow{5}{2cm}{\centering Business} & Pricing & \cmark & \cmark & \cmark & \cmark & \cmark & The pricing information about this product. \\ 

    & User feedback & \cmark & \cmark & \cmark & \cmark & \cmark & User feedback (e.g., rating and comments) of this product.  \\ 

    & Usage statistics &  & \cmark &  &  & \cmark & Number of downloads for this product. \\ 

    \midrule

    \multirow{5}{2cm}{\centering  Product Submission \& Store Review} & Online submission & \cmark &\cmark &  & \cmark & \cmark & Developers can automatically submit new versions of their products online. \\ 

    & Store review policy &  &  &  & \cmark & \cmark & Documentation on policies for developers to follow in order to get approval of the product. \\ 



    \midrule

    \multirow{5}{2cm}{\centering Legal Information} & End user license & \cmark & \cmark & \cmark & \cmark & \cmark &
    Regulations on how users can use this product.  \\ 

    & Developer license &  & \cmark  & \cmark &  &  &
    Regulations on how developers can further expand, integrate, and distribute a product in an authorized way.  \\ 



   \bottomrule

    \end{tabular}


\end{threeparttable}
\end{table*}

\subsection{Comparison among Model Stores}

Among the total of 26 store elements, 20 exist in one or more model stores and  13 are common among the three studied model stores. Below, we detail our comparison results for each dimension:

\begin{itemize}

\item The \textbf{Product Information} dimension contains elements describing the characteristics of the model that is being distributed on the model stores. Only two elements (\texttt{owner} and \texttt{description}) are common among the three model stores. The \texttt{owner} element shows the contact information from the developers who submitted a model, while the \texttt{description} explains its objectives and functionalities. 
In addition to the above elements, ModelDepot and Wolfram provide information regarding the models' \texttt{size} on disk. 
ModelDepot also has a unique \texttt{demo} element allowing users to try out an ML model inside the browser without installing it. 
Models in AWS and Wolfram 
usually include a \texttt{version number} for each release, so that their users can easily tell whether they are using the current version of the model. 

\item The \textbf{Product Documentation} dimension contains the development-specific information related to an ML model. Different from app stores, all three model stores contain \texttt{user instructions}, as the users of ML models generally are organizations/developers. They will likely reuse a model as is in a similar or different product context (transfer learning), re-train a model using the provided training scripts or extend it by adding additional elements to the model. Hence, instead of a purely textual description, the \texttt{user instruction} for ML models generally contains programming examples in the form of scripts (e.g., Jupyter notebooks). 
    
    Different from AWS, ModelDepot and Wolfram contain many models originating from research prototypes or open source software products published on GitHub or authors' websites. The \texttt{origin} information of the models from these two stores is displayed in a dedicated section. So does the information about the \texttt{framework} (e.g., TensorFlow) used to train a model and the \texttt{ML algorithm} (e.g., Convolutional Neural Network). 
  In ModelDepot, the information regarding the \texttt{framework} and the \texttt{ML algorithm} is prominently displayed at the top of each ML model's page. In Wolfram, more detailed \texttt{framework} and \texttt{ML algorithm} information is provided (e.g., number of layers and parameters for neural network architectures). 
  In contrast, only about 4\% of AWS models provide the \texttt{origin} information, 6\% provide \texttt{framework} information, and 20\% provide \texttt{ML algorithm} information. 

Among the three model stores, Wolfram and ModelDepot have dedicated areas to display detailed information about the \texttt{training set} used for a model, and its statistical performance on a test dataset. However, usually only a URL is provided, without deeper discussion of the expected data schema. Furthermore, different performance metrics are used for different models, even for products within the same domain. For example, some image classification models used the overall accuracy metric under 10-fold cross validation, whereas others used ``top-1''/``top-5'' accuracy under 2-fold cross validation. 
Very few ($\sim3\%$) AWS models provide performance results and such information is not presented in a structured manner. Whenever a model is updated to a newer version, it is important to document the changes (e.g., feature updates or bug fixes) in a \texttt{release notes} document. However, such information is missing or poorly presented in model stores. Although AWS contains release notes, they are usually very brief with only one or two sentences. ModelDepot does not contain \texttt{version number} and \texttt{release notes}.



\item The \textbf{Delivery} dimension contains two elements related to the installation and configuration of ML models. Running ML models usually requires specialized hardware (e.g., GPU) or high performance servers. Furthermore, installing and configuring the needed software components for an ML model is a non-trivial task. Hence, all three model stores provide \texttt{deployment instructions}. AWS and Wolfram provide dedicated cloud infrastructure to run all their models, which greatly eases the deployment of these ML models for users. 
While ModelDepot also provides cloud support, it currently only supports one model. 

\item The \textbf{Business} dimension contains three elements related to the business aspects of the products. All three model stores contain \texttt{price} information for their products. This information usually includes the costs of using the store's cloud infrastructure (e.g., VMs and ML APIs). However, the pricing scheme is rather complex and not directly tied to the usage context of end users. For example, AWS charges users on the cloud VM infrastructure and the usage of the model package for training and predicting. Without any performance estimations (e.g., the duration of training/prediction under a particular setup), it is not clear how much one user will be charged for their tasks.
    The \texttt{usage statistics} are missing in AWS and Wolfram. Although ModelDepot provides the number of downloads for each model, it did not provide any information about the types of infrastructure nor the number of API calls for individual products. Such information would be very valuable for software engineers to scale and optimize their ML applications. 

\item The \textbf{Product Submission \& Store Review} dimension contains the information related to submission of a model to the store and to review feedback from the stores. The submission process for AWS and ModelDepot just requires to upload a model online, whereas developers have to contact the store owners of Wolfram in advance to arrange the model submission. None of the three model stores contain any publicly available development policies 
    regarding product reviews and approval. 

\item The \textbf{Legal Information} dimension contains elements related to licensing information of this product.
	For example, all model stores contain \texttt{end user licenses}. The majority of AWS products are developed by commercial companies, whereas the most of the models from Wolfram and ModelDepot are based on research prototypes or open source projects. As such, those models usually adopt open source licences (e.g., Apache or MIT licenses), which allow users to access the models' source code to further modify or extend them.

\end{itemize}



\subsection{Comparison between Model Stores and App Stores}


When comparing the elements between the app and the model stores, we only focus on the elements missing in either all model stores or in both app stores. Under the \textbf{Product Documentation} dimension, there are one unique element in model stores and three unique elements for app stores. 

\begin{itemize}

  \item The \texttt{Demo} element only exists in ModelDepot and is missing in all other model stores and app stores. 
      Compared to mobile apps, which are meant to be downloaded on mobile devices and hence are harder to disable after the expiry of the demo, it is much easier to support model demos as models are meant to be deployed in a container/server. 

  \item Although ML products require access to various computing resources (e.g., images/videos/audio), \texttt{permission}, the list of required computing services (e.g., microphone) or data (e.g., calendar), of models are not explicitly documented.
      The former can be derived through trial-and-error, or by skimming through the annotated scripts. 
      The content of some of the ML products might not be suitable for certain users (a.k.a., \texttt{age} or \texttt{language}). For example, one model in Wolfram is about determining whether an image contains pornographic content.

\end{itemize}

Except \texttt{release notes}, all other elements under the \textbf{Product Documentation} are missing in both app stores. The difference is mainly due to their different target audience (software engineers vs. end users). Apps are products targeted towards the general population, and hence come with a rich 
    GUI. Furthermore, most of these apps provide in-app tutorials when users initially launch them. In contrast, models generally do not come with a GUI but instead correspond to APIs or components that require programming in order to integrate them into an application. Instead of detailed developer documentation, models usually provide sample usage in form of annotated scripts.

The elements under the \textbf{Delivery} dimension are completely disjoint, with two elements only present in model stores, and two only in app stores. Such differences are mainly because the automated product deployment techniques differ between two types of products: apps installed on the users' devices (app stores) or models on the providers' cloud infrastructure (model stores). 
The \texttt{deployment instructions} and \texttt{cloud deployment} information are provided for all model stores, while all app stores check compatibility of apps with the user's device and allow one-click purchase/installation of apps.

Although all elements under the \textbf{Business} dimension exist in both types of stores, the \texttt{pricing} information is presented differently. For model stores, the pricing is usually subscription-based or pay-per-use, whereas mobile apps have a wider range of pricing schemes (e.g., entirely free, one-time purchase, and in-app purchase). 
The \texttt{store review policy} under the \textbf{Submission \& Review} dimension is missing in model stores. As more models are being introduced into the stores, such policies will be needed to protect users and developers. Similar to the \textbf{Product Documentation} dimension above, the \texttt{developer license} element 
is missing under the \textbf{Legal Information} dimension for app stores. 


\subsection{Summary and Implications}


\begin{itemize}

\item \textbf{Emerging Practices}: Since model stores have been introduced recently, only 65\% of the information elements are common across the three model stores. For example, ML model information elements related to technical details like ML algorithms, type of training datasets and cloud deployment, are supported across all three stores. However, some other elements (e.g., demo or release notes) are only present in one or two stores. It would be interesting to study the evolution of the information elements from the model stores, as they are being used by more organizations/developers.

\item  \textbf{Target Audience}: Both model and app stores have several unique elements. For example, model stores contain common usage for each ML model whereas app stores contain age ratings for different apps. This is mainly due to their different target users: app stores for end-users and model stores for organizations/developers.

\item \textbf{Reviewing Policy}: Some important elements in app stores are currently missing in model stores. In particular, there is no clear policy for submitting and reviewing ML models before they can appear in the model stores. The reviewing of ML models is a very challenging task and requires further research in the following three areas: (1) \emph{Requirement specification}: Models used in different context (e.g., health care vs. gaming) need different quality thresholds in order to be usable or safe. For example, how should the safety requirements for radiology-related prediction models be defined?, and (2) \emph{Automated monitoring mechanisms}: To evaluate the safety and the correctness of different ML models under submission, automated monitoring mechanisms are needed; and (3) \emph{Standard quality measurement}: common performance measures of ML models are needed as indictors of qulity of service (QoS) to enable users to compare among similar product offerings.

\item \textbf{Hidden Bias}: Each model in the model store contains three components: the source code, the training dataset, and the trained model(s). However, little information is provided regarding the underlying data distributions and the steps for data pre-processing of the training set. Yet such information is very important in order to identify and remove the hidden bias in models. For example, the deployed ML models can perform poorly, if the images used during training are high resolution images and lower resolution images taken from mobile phones are used in production. Further research is needed to assist organizations/developers to properly identify, report, and remove such bias in model stores in order to yield satisfactory performance of ML applications.

\end{itemize}

\section{RQ2: How unique are the models provided by each model store? }
\label{sec:rq3}

Here we seek to investigate whether different model stores offer their unique offerings of ML models. We first identify the different types of models in model stores, then compare them across model stores.





\subsection{Data Extraction}
\label{sec:extract}

In order to obtain information about all models offered by the three studied model stores, we first developed a model store crawler. Since each model store has a different structure (JSON for AWS, HTML sections for Wolfram and ModelDepot) and displays its data differently (typically using Javascript to dynamically reveal information), we had to write a different crawler for each store leveraging headless Chrome
to obtain the dynamic store content. 
Using the manually labeled information elements used in RQ1, we 
developed parsers to automatically extract the sections of each store. 
In this RQ, we study the most recent snapshot obtained using our crawlers 
at the time of this study (mid March 2019).

\subsection{Quantitative Analysis}
\label{sec:size}

Different model stores have different heuristics to group their models. AWS labels each model using seven criteria (e.g., input and server location) and each model can be under multiple criteria. For example, the input for a computer vision model in AWS can be \emph{image}(s) or \emph{video}(s). The model can be deployed in \emph{US East} or \emph{Europe}. 

After manually studying the grouping criteria of each model store, we decided to group the models based on their input data domain for the following two reasons: (1) it is a common criteria among three stores; and 
(2) each model can only belong to one input domain. Table~\ref{breakdown_using_4_categories} shows the number of models under each group.

AWS has the largest number of models, followed by Wolfram, and ModelDepot. AWS is the only model store with models in all five groups. Neither ModelDepot nor Wolfram contain any models in \texttt{structured data}, while this group contains the majority (45\%) of AWS models. The majority of the ModelDepot (75\%)
and Wolfram (75\%) models are focused on images, which is the second largest group in AWS (27\%). 
All three models stores contain only few models in the \texttt{audio} and \texttt{video} group.

Since RQ1 showed that models in the model stores provide their technical documentations, we 
manually went through each model 
to track their 
\texttt{origin}. As a result, we found that 91\% of the Wolfram and 72\% of the ModelDepot models refer to 34 and 20 academic papers, respectively. One paper/URL may correspond to multiple ML models even in the same store. For example, we found two different models in ModelDepot using the same implementation of one research prototype, but were trained on two different datasets and used in two different contexts: gender recognition and emotion classification. Similar cases also exist in Wolfram. 
Very few (4\%) models in AWS referenced academic papers, each of which corresponds to different AWS models. 
Seven models in Wolfram and three in ModelDepot do not contain paper references but GitHub URLs for the model implementation.
\begin{table}[]
\centering
\caption{The breakdown of ML models under different model stores. Note that AWS contains 231 URLs, each of which corresponds to one model. Yet there are three models that have two URLs for their two different versions. This brings the number of AWS models to 228. We consider two models from two different stores as similar offers, when the ML algorithms, the training dataset(s), and the objective(s) are the same.}
\label{breakdown_using_4_categories}
\begin{tabular}{ccccc}
\toprule

\textbf{Group} & & \textbf{AWS} & \textbf{ModelDepot} & \textbf{Wolfram} \\ 

\midrule

\multirow{2}{*}{Image} & Count & 61 (27\%) & 24 (75\%) & 59 (75\%) \\ 
& Similar & 2 (0.8\%) & 6 (19\%) & 7 (9\%) \\

\midrule

\multirow{2}{*}{Video} & Count & 13 (6\%) & 2 (6\%) & 0 (0\%) \\ 
& Similar & - & - & - \\

\midrule

Natural  & Count & 35 (15\%) & 5 (16\%) & 18 (23\%) \\ 
language & Similar & - & 1 (3\%) & 1 (1\%) \\

\midrule

\multirow{2}{*}{Audio} & Count & 12 (5\%) & 1 (3\%) & 2 (2\%) \\ 
& Similar & - & - & - \\

\midrule

\multirow{2}{*}{Structured} & Count & 107 (47\%) & - & - \\ 
& Similar & - & - & - \\

\midrule

\multirow{2}{*}{Total} & Count & 228 (100\%) & 32 (100\%) & 79 (100\%) \\ 
& Similar & 2 (0.8\%) & 7 (22\%) & 8 (10\%) \\

\bottomrule

\end{tabular}
\end{table}

\subsection{Similar Offerings of ML Models}
\label{sec:overlap}

We consider two ML models from different model stores as similar offerings (a.k.a., similar models), if they share three information elements in common: ML algorithms, training datasets, and objectives. For example, all three model stores contain an image classification model that uses the same algorithm (ResNet50) and training dataset (Imagenet). Most ML models have such information elements in their individual product webpage. Note that similar ML models may not be exactly identical. For example, although two similar ML models use the same ML algorithms, their underlying implementations can be different.
Table~\ref{breakdown_using_4_categories} shows the results. there are only two similar models in AWS to the other two stores, whereas ModelDepot and Wolfram had seven and eight common models. Most of the similar ML models were found under the \textbf{image} group.

\subsection{Summary and Implications}


\begin{itemize}

  \item \textbf{Product Maturity}: More than 70\% of the models from ModelDepot and Wolfram are based on research prototypes. This demonstrates the practical impact of current AI research, which can be converted into production-ready models in a relatively short time-frame. It would to be interesting to track their future development activities of such models to understand the unique challenges and opportunities for maintaining and evolving ML models.

 \item \textbf{Cross-store Support}: The amount of similar models across different model stores is very small. This is mainly due to vendor lock-in. Migrating one model to different stores requires adapting it to different frameworks, like SageMaker for AWS and the Wolfram language for Wolfram. Similar to mobile app stores, cross-platform frameworks for developing and maintaining ML models are needed.

\end{itemize}

\section{Conclusions and Future Work}
\label{sec:conclusionAndFutureWork}

This paper presented an exploratory study on ML model stores. We first empirically compared the information elements among three model stores and two app stores. Since model stores have been introduced fairly recently, only 65\% of the elements are common among model stores. We found some elements (e.g., cloud deployment and user instructions) which are unique to the model stores. Certain elements (e.g., review policy) which are present in app stores are missing in model stores. Further studies of the models inside the three model stores showed very few offerings of the similar ML models among the model stores, with the majority of the ML models from Wolfram and ModelDepot originating from research prototypes. In the future, better support for effective reviewing of ML models in terms of safety and quality are needed in model stores. Before integrating into ML applications, automated methods are needed to detect, report, and remove hidden bias in pre-trained ML models.



\balance

\bibliographystyle{abbrv}
\bibliography{xmkBibLib}

\end{document}